\documentclass[twocolumn]{aastex62}
\usepackage{amsmath,amssymb}
\usepackage{graphicx}
\usepackage{hyperref}
\begin{document}

\title{Diagnosing the Optically Thick/Thin Features Using the Intensity Ratio of \ion{Si}{4} Resonance Lines in Solar Flares}
\author{Yi-An Zhou}
\affiliation{School of Astronomy and Space Science, Nanjing University, Nanjing 210023, People's Republic of China}
\affiliation{Key Laboratory for Modern Astronomy and Astrophysics (Nanjing University), Ministry of Education, Nanjing 210023, People's Republic of China}
\author{Jie Hong}
\affiliation{School of Astronomy and Space Science, Nanjing University, Nanjing 210023, People's Republic of China}
\affiliation{Key Laboratory for Modern Astronomy and Astrophysics (Nanjing University), Ministry of Education, Nanjing 210023, People's Republic of China}
\author{Y. Li}
\affiliation{Key Laboratory of Dark Matter and Space Astronomy, Purple Mountain Observatory, Chinese Academy of Sciences, Nanjing 210033, People's Republic of China}
\author{M. D. Ding}
\affiliation{School of Astronomy and Space Science, Nanjing University, Nanjing 210023, People's Republic of China}
\affiliation{Key Laboratory for Modern Astronomy and Astrophysics (Nanjing University), Ministry of Education, Nanjing 210023, People's Republic of China}

\email{jiehong@nju.edu.cn, dmd@nju.edu.cn}

\begin{abstract}
In the optically thin regime, the intensity ratio of the two \ion{Si}{4} resonance lines (1394 and 1403 \AA) are  {theoretically} the same as  {the ratio of} their oscillator strengths, which is exactly 2. 
Here, we study the ratio of  {the integrated} intensity of the \ion{Si}{4} lines  ($R=\int I_{1394}(\lambda)\mathrm{d}\lambda/\int I_{1403}(\lambda)\mathrm{d}\lambda$) and the ratio of intensity at each wavelength point ($r(\Delta\lambda)=I_{1394}(\Delta\lambda)/I_{1403}(\Delta\lambda)$)  {in} two solar flares observed by the Interface Region Imaging Spectrograph. 
We find that at flare ribbons, the ratio $R$  {ranges} from 1.8 to 2.3 and would generally decrease when the ribbons sweep across the slit position. Besides, the distribution of $r(\Delta\lambda)$ shows a descending trend from the blue wing to the red wing. 
In loop cases, the \ion{Si}{4} line presents a wide profile with a central reversal. The ratio $R$ deviates little from 2, but the ratio $r(\Delta\lambda)$ can  {vary from} 1.3 near the line center  {to} greater than 2 in the line wings.
Hence we conclude that in flare conditions, the ratio $r(\Delta\lambda)$  {varies across the line}, due to the variation of the opacity at the line center and line wings. 
 {We notice that, although} the ratio $r(\Delta\lambda)$ could present a value which deviates from 2 as a result of the opacity effect near the line center, the ratio $R$ is  {still} close to 2.  {Therefore, caution should be taken when using} the ratio of  {the} integrated intensity  {of the Si IV lines to diagnose} the opacity effect. 
\end{abstract}


\section{Introduction}
\label{sec-intro}
Solar flares can release a large amount of energy within a short period of time and affect all layers of the solar atmosphere including the chromosphere, the transition region (TR) and the corona \citep{Fletcher2011}. 
With a formation temperature of $\sim$10$^{4.8}$ K, the two cool \ion{Si}{4} resonance lines (1394 and 1403 \AA) are excellent candidates to directly reveal the  {physical} properties of the TR  {when small scale events occur, }such as  {transition region explosive events} \citep{Innes1997, Teriaca2004, Huang2014, Huang2015, Huang2017, Chen2019, Reep2016, Warren2016}, transient flows \citep{Mcintosh2009, Kleint2014}, {UV bursts} \citep{Peter2014,  Hong2021},  {spicules} \citep{Tian2014b},  {microflares and nanoflares} \citep{Testa2014, Polito2018, Hannah2019}.  {They have also been used in the study of solar flares, especially the properties of }chromospheric evaporation \citep{Tian2014a, Li2015, LiD2017,Li2017, Li2019, Brosius2016, Brosius2018} and chromospheric condensation \citep{Zhang2016, Yu2020}.

In the quiet Sun, the density of the TR and corona is low enough that the emitted \ion{Si}{4} photons can escape without distinct absorption, scattering or reemission. Thus, these lines are usually regarded as optically thin. In the optically thin regime, the intensity ratio of these two lines, should be the same as the ratio of their oscillator strengths, which is exactly  {2} \citep{Math1999}.  This theoretical value can also be obtained with the CHIANTI database \citep{Dere1997, delZanna2021}. 

However, the intensity ratio can deviate from  {2} in many observations. \cite{Yan2015} revealed self-absorption features of the \ion{Si}{4} line during a transient brightening event in an active region (AR). The line ratio of \ion{Si}{4} lines falls from $\sim$2.0 to $\sim$1.7  {with the occurrence of the event }and increases back to $\sim$2.0 near the end of the event. This self-absorption feature is interpreted as a result of the overlying cool TR loops.
\cite{Gon2018} also  {presented} an observation that the  intensity ratio between the \ion{Si}{4} 1394 and 1403 \AA\ lines is greater than  {2}. This deviation is rendered to be the result of resonant scattering. 
Besides, \cite{Tri2020}  {made} a detailed study of the  {spatial distribution and time evolution} of the intensity ratio between the two \ion{Si}{4} lines in an emerging flux region. The intensity ratio varies a lot from the early phase to the late phase, with different values at the periphery and the core of an AR. The ratio in the quiet Sun regions is generally close to  {2}, and the difference of the ratio between the AR and the quiet Sun  {is usually ascribed} to the opacity effect. 
In addition, \cite{Mulay2021}  {found} that the \ion{Si}{4} lines are mostly optically thin when the ratio is $\sim$2 at locations with strong H$_2$ emissions during the impulsive phase of  {a} flare.  {However,} at some locations of flare ribbons,  {the} increased opacity causes the ratio to deviate from  {2}. 
Recently, a detailed theoretical calculation of the \ion{Si}{4} line profiles  {in} flare  {models confirmed} that due to the opacity effect, the intensity ratio of the \ion{Si}{4} lines at flare ribbons could {vary} from 1.8 to 2.3 \citep{Kerr2019}.
 {Given the diverse results in observations and numerical simulations, it is still unclear if the Si IV lines are formed in optically thin or thick conditions in various solar activities, and how reliable it is to use the intensity ratio of the two lines to diagnose the opacity effects. }

As for the intensity ratio of the \ion{Si}{4} lines, previous studies only  {focused on} the ratio of the total intensity integrated over wavelength. However, in flare conditions  {the opacity should significantly vary from} the \ion{Si}{4} line center  {to} line wings. The photons  {at line wings} can still escape freely  {from the solar surface} while  {those at} the line center might not \citep{Kerr2019}. In this case, it would be inappropriate to use the ratio of the  {integrated} intensity as a diagnostics of opacity, since the opacity at the line center  {is} quite different  {from that at line wings}.

In this paper, we focus on the \ion{Si}{4} line profiles and  {investigate} (1) the ratio of the  {integrated} intensity of the \ion{Si}{4} lines ($R=\int I_{1394}(\lambda)\mathrm{d}\lambda/\int I_{1403}(\lambda)\mathrm{d}\lambda$) and (2) the ratio of the intensity at each wavelength  {point relative to the line center} of the \ion{Si}{4} lines ($r(\Delta\lambda)=I_{1394}(\Delta\lambda)/I_{1403}(\Delta\lambda)$). 
 {The latter is based on the consideration that the two Si IV lines are formed at similar heights and thus their Doppler widths are almost the same in most cases.} 
The rest of this paper is organized as follows. The observations from the Interface Region Imaging Spectrograph (IRIS; \cite{De14}) and data reduction are  {described} in Section \ref{sec-obs-data}. The detailed results of the \ion{Si}{4} line profiles and line ratios are shown in Section \ref{sec-res}. 
Finally, Section \ref{sec-con-discuz} presents a summary and discussions.

\section{Observations and Data Reduction}
\label{sec-obs-data}

\begin{figure*}
	\centering
	\includegraphics[width=\linewidth]{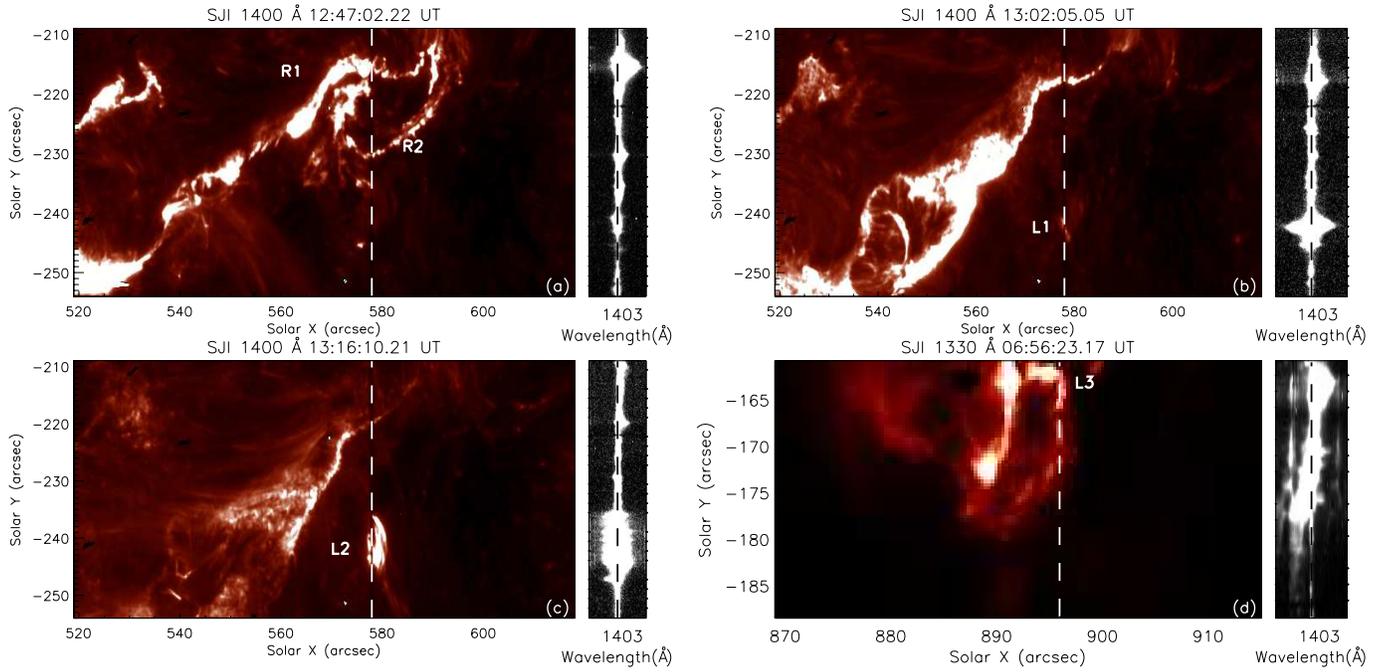}
	\caption{
IRIS SJIs and \ion{Si}{4} 1403 \AA\ spectra showing an overview of two solar flares. 
Panels (a)-(c) are for the M7.3 flare and (d) for the C1.7 flare. 
Flare ribbons (R1 and R2) and loops (L1--L3) are marked in each panel. White dashed lines in the SJIs mark the slit position, and black dashed lines in the spectra mark the line center of \ion{Si}{4} 1403 \AA.}
    \label{fig_sji}
\end{figure*}

We study two solar flares  {(with M7.3 and C1.7 classes)} observed with IRIS. The IRIS performed high-cadence sit-and-stare spectral observations with a slit width of 0.\arcsec33 and a time cadence of 9.8 s for both flares. The spectral resolution of the FUV band covering the \ion{Si}{4} lines is $\sim$0.025 \AA\ . An overview of these two flares with IRIS slit-jaw images (SJIs) as well as the spectra of the \ion{Si}{4} line at 1403 \AA\  is displayed in Figure \ref{fig_sji}.

The M7.3 flare under study started at $\sim$12:31 UT, peaked at $\sim$13:03 UT and ended at $\sim$13:20 UT on 2014 April 18. The IRIS data cover the time period from $\sim$12:33 UT to $\sim$17:18 UT. In the impulsive phase of the flare, one can see clearly two flare ribbons {(Figure \ref{fig_sji}{(a)}, marked by R1 and R2, respectively)} that lasted for minutes. The northern ribbon (R1) disappeared at $\sim$ 13:20 UT, while the southern ribbon (R2) disappeared earlier at $\sim$12:57 UT. We notice a hook structure of R2, which firstly appeared at $\sim$12:43 UT, reached the peak intensity at $\sim$12:47 UT (Figure \ref{fig_sji}{(a)}) and disappeared at $\sim$12:50 UT. The \ion{Si}{4} spectra at these flare ribbons display strong line and continuum emissions. There also appeared loop-like structures to the south of the flare ribbons which became more obvious at later time ($\sim$13:02 UT and {$\sim$13:16 UT}, Figure \ref{fig_sji}{(b)} and \ref{fig_sji}{(c)}, marked by L1 and L2, respectively). The \ion{Si}{4} spectra at these loops display very wide line profiles.
 As for the C1.7 flare that \cite{Zhou2020} studied,  it occurred in the active region NOAA 12673 near the solar limb. This flare started at $\sim$06:51 UT and peaked at $\sim$06:56 UT on 2017 September 9. Figure \ref{fig_sji}{(d)} illustrates the loop-like structures (L3) {at  $\sim$06:56 UT. }

We then calculate the intensity ratio of the \ion{Si}{4} lines. As described above, we calculate both the ratio of wavelength-integrated intensity, $R$, and the ratio of intensity at each wavelength point, $r(\Delta\lambda)$. To improve the reliability, we first subtract the background continuum and the blended lines before calculations. The wavelength range for the calculations of both $R$ and $r(\Delta\lambda)$ is then specifically chosen to be where the intensity is three times larger than the standard deviation of the intensity fluctuations in the far wings. The data points where the line profiles are saturated or influenced by background cosmic rays are discarded.

\section{Results}
\label{sec-res}

\begin{figure*}
	\centering
	\includegraphics[width=\linewidth]{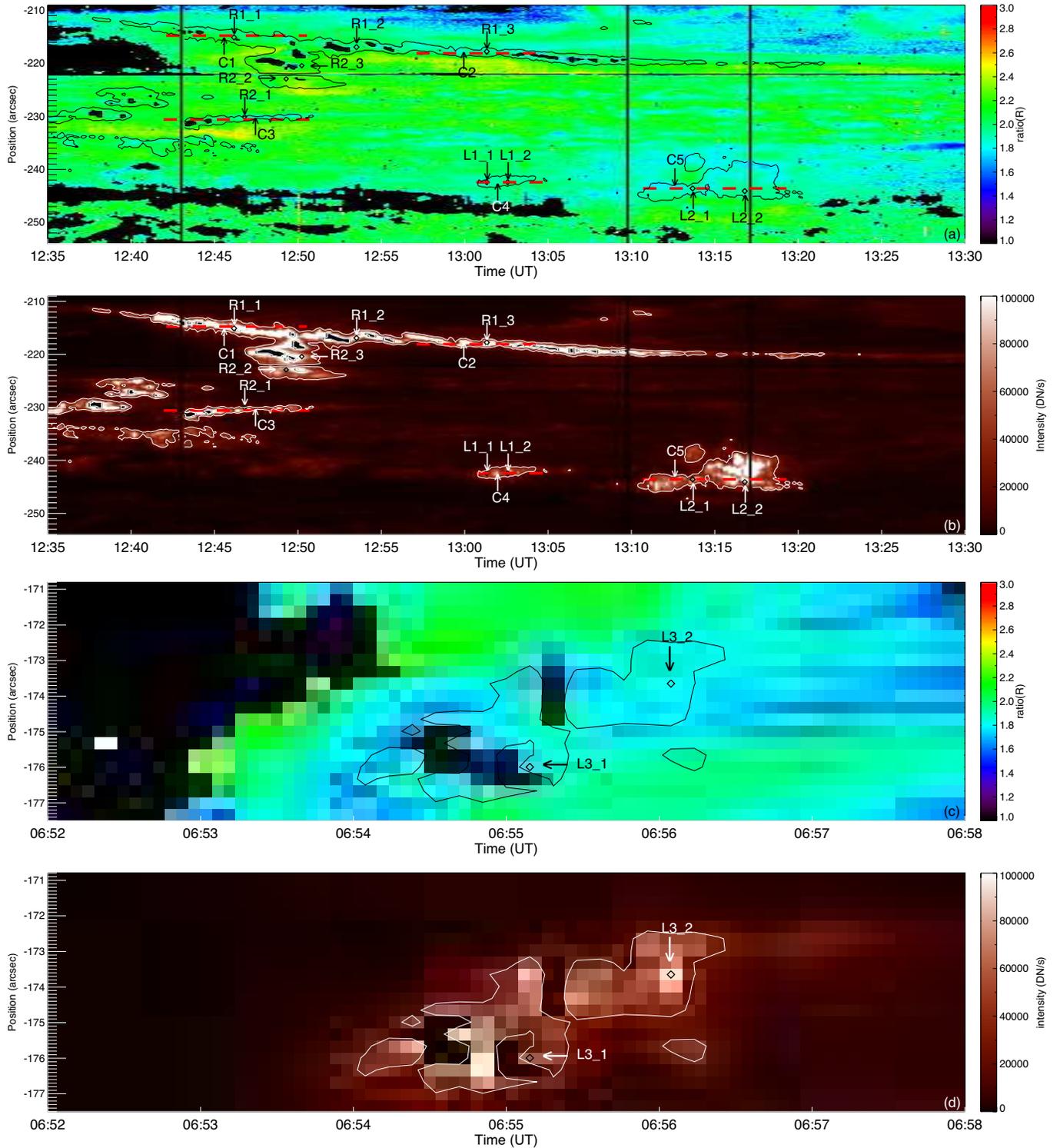}
	\caption{
Panels (a)--(b) : The space-time maps of the intensity ratio $R$ and the total intensity of the \ion{Si}{4} 1394 \AA\ line for the M7.3 solar flare. 
Specific positions and horizontal cuts for further study are marked with black diamonds and red dashed lines, respectively. 
The contours indicate an intensity level of three times the average value of the total intensity of the \ion{Si}{4} 1394 \AA\ line over the quiet region.
The black area refers to positions where the spectra were overexposed or with a low signal-to-noise ratio, which are not used for analysis.
Panels (c)--(d) : Same as panels (a)--(b), but for the C1.7 flare.
	}
    \label{ratio_map} 
\end{figure*}

The upper two panels of Figure \ref{ratio_map} display space-time maps of the ratio $R$ between the \ion{Si}{4} 1394 and 1403 \AA\ lines and the integrated \ion{Si}{4} 1394 \AA\ intensity at slit positions for the M7.3 flare, while the lower two panels are for the C1.7 flare. The intensity contours in these panels show the evolutions of the flare ribbons and loops as described in Figure \ref{fig_sji}.
We choose several positions in the flare ribbons and loops (labeled Rx\_x and Lx\_x) for further study, marked with black diamonds. 
The line profiles of these typical points are shown in Figures \ref{ratio_ribbon} and \ref{ratio_loop}.
We also choose five different slit positions and make horizontal cuts (labeled C1 to C5) in the space-time map to study the time evolution, marked with {red} dashed lines. The time evolution of the ratio $R$ and line profiles at selected  {time instants} are plotted in Figure \ref{ratio_evo}.

\subsection{{Intensity ratio at flare ribbons}}
\label{sec-ribbon}
\begin{figure*}
	\centering
	\includegraphics[width=\linewidth]{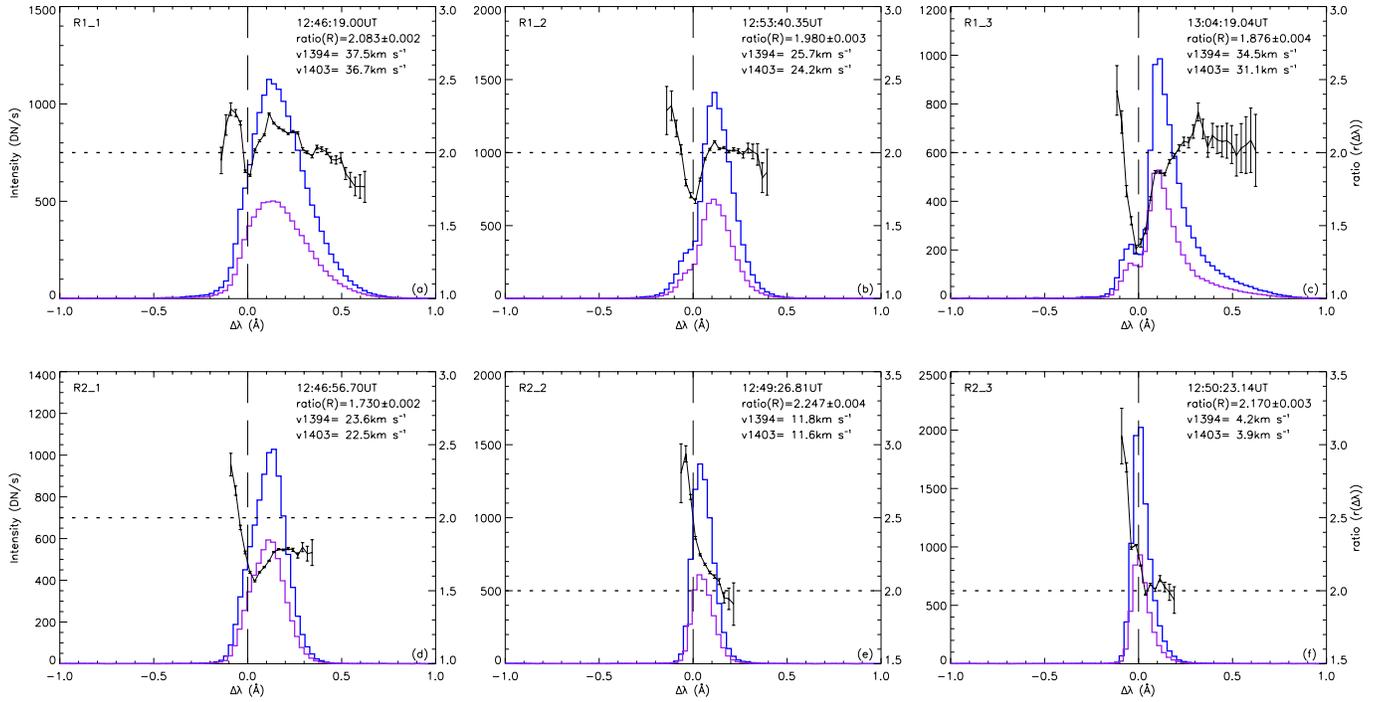}
	\caption{
Line profiles of \ion{Si}{4} 1394 \AA\ (blue) and 1403 \AA\ (purple) at ribbon positions. 
In each panel, the dashed vertical line denotes the line center of the two \ion{Si}{4} lines, and the dashed horizontal line marks $r(\Delta\lambda)=2$. 
The lines with {error bars} show the ratio ($r(\Delta\lambda)$) between the two lines. The velocities in each panel are derived from the centroid of the \ion{Si}{4} line profiles.}
\label{ratio_ribbon}
\end{figure*}

As shown in Figure \ref{ratio_map}, the ratio $R$ is mostly close to 2 in the flare ribbons R1 and R2, but with deviations at some specific positions \citep{Mulay2021}.
The line profiles of \ion{Si}{4} 1394 and 1403 \AA\ at the positions marked by black diamonds along flare ribbons are plotted in Figure \ref{ratio_ribbon} in blue and purple, respectively. {The ratio $r(\Delta\lambda)$ at each wavelength point is overplotted with error bars.}
In this M7.3 flare, it is seen that most of the \ion{Si}{4} lines present red asymmetries or red-shifted profiles in the ribbon area. For profiles {at R1\_1 and R1\_2 (Figure \ref{ratio_ribbon}{(a)--(b)})} , though the ratio $R$ deviates little from 2, the ratio $r(\Delta\lambda)$ presents a tendency of decrease from the blue wing to the red wing, {with a dip near the line center}. The ratio $r(\Delta\lambda)$ can be larger than 2 in the blue wing while smaller than 2 in the red wing. 
The profiles at R1\_3 present double peaks (Figure \ref{ratio_ribbon}{(c)}). {At this position}, the ratio $R$ is $\sim$1.8. The ratio $r(\Delta\lambda)$ shows a similar distribution along wavelength to those at R1\_1 and R1\_2. We also find that the dip at R1\_3 is much obvious than others.
The variation of the ratio $r(\Delta\lambda)$ along wavelength at R2 is generally similar to that at R1. However, the {ratio $R$ at R2\_1 is much smaller than 2, while that at R2\_2 and R2\_3 is larger than 2.}

\subsection{{Intensity ratio at loop cases}}
\label{sec-loop}

Figure \ref{ratio_map} shows that the ratio $R$ at the loop structures L1 and L2 are mostly close to 2, while at L3 it obviously deviates from 2.

\begin{figure*}
	\centering
	\includegraphics[width=\linewidth]{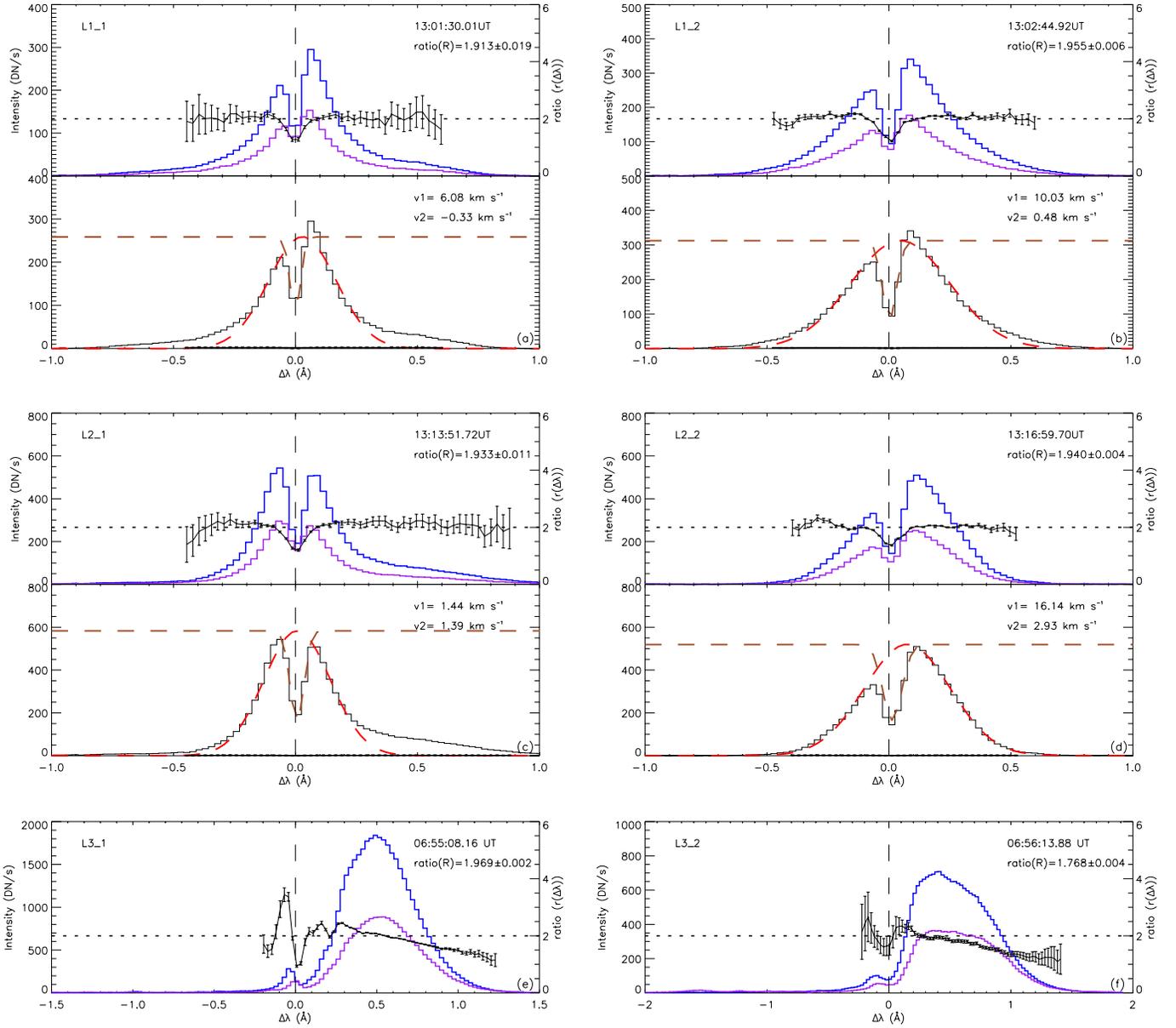}
	\caption{
Same as Figure~\ref{ratio_ribbon}, but for loop positions. The two velocities $v_{1}$ and $v_{2}$ in panels (a)--(d) are derived from Gaussian fittings of the line wings and the central reversal, respectively.}
\label{ratio_loop}
\end{figure*}

Figure \ref{ratio_loop}{(a)--(d)} presents the line profiles and intensity ratios of the \ion{Si}{4} 1394 and 1403 \AA\ lines at the positions marked by black diamonds of the M7.3 flare. The most remarkable feature of these line profiles is an increased width with a central reversal. The central reversal has a depth of about half the peak intensity. Thus, we fit the line wings and the central reversal of the  \ion{Si}{4} 1394 \AA\ line with two Gaussian functions to derive the Doppler velocities corresponding to the plasma motions in the formation heights of line wings and line center, respectively. The central reversal is {nearly} static, while the line wings shows a small red asymmetry. In Figure \ref{ratio_loop}{(c)}, it is seen that the the line emission at red wing is obviously enhanced, which may reflect the existence of some  downward-moving plasma with large velocities.
 
In addition, the wavelength distribution of the ratio $r(\Delta\lambda)$ is different from those in the flare ribbons. The ratio $r(\Delta\lambda)$ here shows a symmetric distribution in the wings while the dip at line center is still very obvious. 

Figure \ref{ratio_loop}{(e)} and \ref{ratio_loop}{(f)} illustrates the line profiles and intensity ratios of the \ion{Si}{4} 1394 and 1403 \AA\ lines at L3\_1 and L3\_2 for the C1.7 flare.  
{For the L3 loop, there appear notable continuum emissions at these positions ({Figure \ref{fig_sji}}), reflecting a possible overlapping of the flare loops with the flare ribbons along the line of sight.}
These line profiles show a very strong red peak, which was suggested to result from reconnection downflows \citep{Zhou2020}. The wavelength distribution of the ratio $r(\Delta\lambda)$ looks similar to those of R1, with a general decrease from the blue wing to the red wing, and a dip at line center. 

\subsection{Time evolution of intensity ratio}
\label{time-evo}
The time evolution of the ratio $R$ and the {integrated intensity of \ion{Si}{4} 1394 \AA} at different slit positions as well as some selected \ion{Si}{4} 1394 \AA\ line profiles are presented in Figure \ref{ratio_evo}. 
It is seen that when the flare ribbons R1 and R2 sweep across the slit position, the {integrated} line intensity increases, while the ratio $R$ decreases correspondingly, as shown for ribbon cases C1 to C3. The value of ratio $R$ is generally below 2 at flare ribbons but over 2 at quiet Sun. However, we notice that the ratio $R$ at C2 could {quickly} rise back to 2 after an initial fall. This is because at this time, the line profiles are very wide and show a strong peak in the red wing. The contribution from the line wing, which is closer to the optically thin scenario, thus increases the values of $R$.  
For loop cases C4 and C5, the intensity and ratio $R$ do not vary much with time. The right panels of Figure \ref{ratio_evo} also display distinct dips at the center of \ion{Si}{4} 1394 \AA\ line profiles when loop structures appear at the slit positions, but the ratio $R$ is still close to 2.

\section{Conclusions and Discussions}
\label{sec-con-discuz}

In this paper, we carry out an observational study about the intensity ratio between the \ion{Si}{4} 1394 and 1403 \AA\ lines in two solar flares.  We calculate the ratio of the integrated intensity, $R$, and the ratio of line intensity at each wavelength point, $r(\Delta\lambda)$, and explore their relation with the line opacity.
Our observational results are summarized as follows:
\begin{itemize}
\item At the flare ribbons (R1 and R2), most of the \ion{Si}{4} line profiles show obvious redshifts or red asymmetries with the ratio $R$ ranging from {1.8} to 2.3. When the flare ribbons sweep across the slit position, the ratio $R$ would generally decrease. The wavelength distribution of $r(\Delta\lambda)$ shows a descending trend from the blue wing to the red wing, with a dip in the line center, in particular at R1. 
\item At the loop structures (L1 and L2), the  \ion{Si}{4} line presents a wide profile with a central reversal. The line wings show a small redshift, while the central reversal is nearly static. The ratio $R$ deviates little from 2.0, while the ratio $r(\Delta\lambda)$ varies in a wide range, which is only {1.3} at the line center. Nevertheless, the distribution of $r(\Delta\lambda)$ is nearly symmetric  with wavelength.
\item At some locations where the flare loops and flare ribbons possibly overlap (L3), the ratio $r(\Delta\lambda)$ displays a similar wavelength distribution to that of R1.
\end{itemize}

The interesting result is that the ratio $r(\Delta\lambda)$ can vary with wavelength in the line. This is reasonable given the fact that the opacity at the line center and line wings could be quite different under flare conditions. In fact, modeling of the \ion{Si}{4} lines at flare ribbons shows that for an intermediate flare, the opacity effect would not be negligible \citep{Kerr2019}. The opacity effect could explain the wavelength distribution of $r(\Delta\lambda)$, especially the central dip, at R1, R2 {and} L3. {We notice that when the line profiles are significantly redshifted (L3\_2), the ratio $r(\Delta\lambda)$ at the redshifted wing would be smaller than that in the line core. This suggests that the maximum opacity at each height always coincides with the local fluid velocity, as reported from previous flare simulations \citep{Carlsson1997,Kuridze2015,Kerr2019}.}

Under flare conditions, the values of $R$ can also vary from case to case. For example, it can be close to 2 at R1\_1 and R1\_2, less than 2 at R1\_3 and R2\_1, {and greater than 2 at R2\_2 and R2\_3. The opacity effect cannot explain the cases where $R$ is greater than 2. One possible explanation for this result is the decreased contribution of the collisional excitations when resonant scattering is dominant, as proposed by \cite{Gon2018}, who reported a negative relation between line ratio and electron density. In this work, we have estimated the electron density from the ratio of \ion{O}{4} 1401.2 and 1399.8 \AA\ lines. We find that the electron density is around 10$^{10.3}$ cm$^{-3}$ in regions where $R>2$, and is around 10$^{11.4}$ cm$^{-3}$ in regions where $R<2$. These results agree with that of \cite{Gon2018}. 
\cite{Rose2008} proposed that the specific geometry of the flaring plasma could also be a factor making the line ratio greater than 2. 
Moreover, the formation heights of the two lines are slightly different. Comparatively, the 1394 \AA\ line is formed slightly higher than the 1403 \AA\ line. If the line source function has a larger gradient  between the formation heights during the impulsive heating, it could also lead to a line ratio greater than 2.
More observations and radiative hydrodynamic simulations are required to investigate the detailed physical mechanism that leads to these increased line ratios.}

\begin{figure*}
	\centering
	\includegraphics[width=\linewidth]{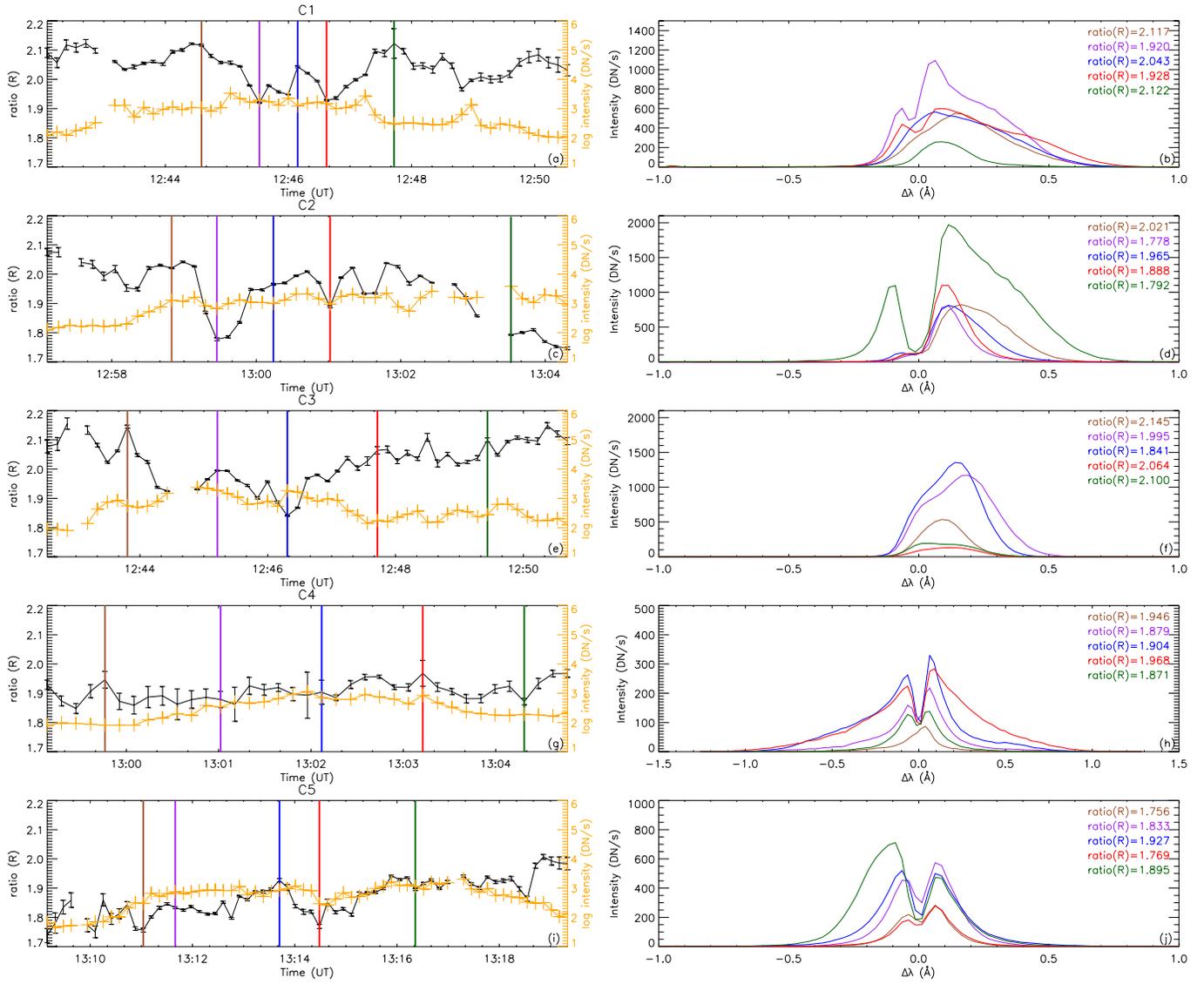}
	\caption{
Left panels : Time evolution of the ratio $R$ {(black curves with error bars)} and total intensity of the \ion{Si}{4} 1394 \AA\ (plus signs) at different slit positions.
Right panels: The selected \ion{Si}{4} 1394 \AA\ line profiles at the slit positions shown in the left panels.}
    \label{ratio_evo}
\end{figure*}

The central reversal in the Si IV line profiles in L1 and L2 looks very similar to the results of \citet{Yan2015}, {which was explained as an opacity effect.} Correspondingly, the value of $r(\Delta\lambda)$ can reach a central dip as low as 1.3, which is a typical feature of an optically thick case. However, the ratio $R$ is still close to 2, as seen from the time evolution of this value (Figure \ref{ratio_evo}).
{In these cases, the photons in the optically thick line core could be scattered to the wide line wings and easily escape since the line wings are still optically thin. 
Therefore, although the wavelength dependent line ratio $r(\Delta\lambda)$ can significantly vary across the line, if the line ratio $R$ of the integrated intensity is still equal to 2, it still implies that the whole line can be regarded as effectively optically thin.
}

Hence, we argue that the ratio of wavelength-integrated intensity cannot be served as a sole indicator of the opacity effect, especially when the line profiles are very wide. 
The line center and line wings could have different opacity effects, to say, the line center becomes optically thick while the line wings still keep optically thin in flare cases. Since the contribution of line wings is usually over that of the line center, the ratio of integrated intensity seems insensitive to the flare conditions.
Therefore, it is highly required to use the ratio of intensity at each wavelength point to investigate the opacity effects of the \ion{Si}{4} lines, in addition to careful checking of the line profiles. 
We also note that, although the two \ion{Si}{4} lines are formed at close heights, they could still suffer from different Doppler shifts or asymmetries in the case that the flare-induced plasma velocity changes steeply (like chromopsheric condensation). This can explain why the distribution of $r(\Delta\lambda)$ is sometimes asymmetric, in particular at flare ribbons. How the asymmetry of $r(\Delta\lambda)$ is related to flare dynamics deserves further investigation. 

\acknowledgments
This work was supported by National Key R\&D Program of China under grant 2021YFA1600504, and by NSFC under grants 11903020, 11733003, 12127901, 11873095, and 11961131002. Y.L. is supported by the CAS Pioneer Talents Program for Young Scientists and XDA15052200, XDA15320301, and XDA15320103. IRIS is a NASA small explorer mission developed and operated by LMSAL with mission operations executed at the NASA Ames Research center and major contributions to downlink communications funded by the Norwegian Space Center (NSC, Norway) through an ESA PRODEX contract.

\clearpage

\end{document}